# A Generalized Recursive Algorithm for Binary Multiplication based on Vedic Mathematics

Ajinkya Kale, Shaunak Vaidya, Ashish Joglekar,
kaleajinkya@gmail.com, vaidya.shaunak@gmail.com, avjcoep@gmail.com

***Abstract*- A generalized algorithm for multiplication is proposed through recursive application of the Nikhilam Sutra from Vedic Mathematics, operating in radix - 2 number system environment suitable for digital platforms. Statistical analysis has been carried out based on the number of recursions profile as a function of the smaller multiplicand. The proposed algorithm is efficient for smaller multiplicands as well, unlike most of the asymptotically fast algorithms. Further, a basic block schematic of Hardware Implementation of our algorithm is suggested to exploit parallelism and speed up the implementation of the algorithm in a multiprocessor environment.**

***Keywords*- Vedic Mathematics, Algorithm, Time Complexity, Multiplication, Recursion, Hardware, Computing**

## I. Introduction

*Vedic Mathematics*:

- Vedic Mathematics deals mainly with 16 Sūtras and their applications for carrying out tedious and cumbersome arithmetical operations, and to a very large extent, executing them mentally; see[5].
- *Nikhilam Navataścaramam Daśatah Sūtra (*or simply *Nikhilam Sūtra)* is one of these 16 Sūtras used for multiplication and has been successfully applied to overcome drawbacks of conventional schemes.

This is where the fields of modern computing and Vedic Mathematics converge.

*Need For Efficient Multiplication Algorithm*:

- Use of numerical methods is prevalent in most software algorithms. Such applications demand an efficient code for basic mathematical operations, one of them being multiplication.
- Real Time Systems demand instantaneous response to environmental variables and quick execution of taken decision.
- Multiplication algorithms find applications in Digital Signal Processing (DSP) for discrete Fourier transforms, Fast Fourier transforms, convolution, digital filters, etc. Therefore any new multiplication algorithm opens up a new approach for improving existing schemes.

This calls for a 'time efficient' algorithm for 'multiplication' to improve processor throughput.

*Existing Algorithms*:

- The naive binary multiplication algorithm has a time complexity O(n^2) where n is the number of bits of the numbers being multiplied.
- The Schönhage-Strassen algorithm is an asymptotically fast multiplication algorithm for large integers. It was developed by Arnold Schönhage and Volker Strassen in 1971. The run-time bit complexity is, in Big O notation, O(n log n (log log n)) , while the arithmetic complexity is O(n log n). The algorithm uses recursive Fast Fourier transforms in rings with elements, a specific type of number theoretic transform.
- The Schönhage-Strassen algorithm was the asymptotically fastest multiplication method known from 1971 to 2007 when a new method, Fürer's algorithm, was announced with lower asymptotic complexity.
- The drawback of the Schönhage-Strassen algorithm is that it starts to outperform older methods such as Karatsuba and Toom–Cook multiplication for numbers beyond 10,000 to 40,000 decimal digits and so is the case with most of the asymptotically fast algorithms. Hence these algorithms are not suitable for general purpose digital systems.
- There is no algorithm in our knowledge whose complexity depends upon ratio of 0s and 1s in multiplicands rather than magnitude of multiplicand.

This motivated us to come up with an algorithm which is 'time efficient' over the entire range of numbers for more generalized applications.

## II. Explanation Of Nikhilam Sūtra

One of the 16 Sūtras of Vedic Mathematics, Nikhilam Sūtra stated algebraically as follows:

Consider two numbers $n_1$ and $n_2$ such that
$n_1 = (x - a) \quad a = (x - n_1)$
$n_2 = (x - b) \quad b = (x - n_2)$
where $x =$ base, $a, b =$ differences from base, then

$$n_1 \times n_2 = (x - a)(x - b) = x((x - a) + (x - b) - x) + a * b$$

Thus the Nikhilam Sūtra effectively breaks up a large multiplication $n_1 \times n_2$ into a small multiplication $(a \times b)$ and addition $[(x - a) + (x - b) - x]$.

*Case 1.1:* Base as power of 10, $n_1$ and $n_2$ less than base:
Consider $n_1 = 99$ and $n_2 = 98$

EXAMPLE 1.1

| **Number** | **Difference** | **Base $x$** |
|---|---|---|
| $(x - a)$<br>99 | $a$<br>+01 | 100 |
| $(x - b)$<br>98 | $b$<br>+02 | |
| $(x - a) + (x - b) - x$<br>**97** | $ab$<br>**02** | |

Hence answer $= 9702$

*Case 1.2:* Base as power of 10, $n_1$ and $n_2$ greater than base:
Consider $n_1 = 102$ and $n_2 = 101$

EXAMPLE 1.2

| **Number** | **Difference** | **Base $x$** |
|---|---|---|
| $(x - a)$<br>102 | $a$<br>-02 | 100 |
| $(x - b)$<br>101 | $b$ -<br>01 | |
| $(x - a) + (x - b) - x$<br>**103** | $ab$<br>**02** | |

Hence answer $= 10302$

*Case 1.3:* Base as power of 10, base between $n_1$ and $n_2$:
Consider $n_1 = 101$ and $n_2 = 99$

EXAMPLE 1.3

| **Number** | **Difference** | **Base $x$** |
|---|---|---|
| $(x - a)$<br>101 | $a$<br>-01 | 100 |
| $(x - b)$<br>99 | $b$<br>+01 | |
| $(x - a) + (x - b) - x$<br>**100** | $ab$<br>**-01** | |

The product *ab* being negative is to be subtracted from 10000. Hence answer $= 9999$

*Case 2:* Modified Base, Multiplication Method:
Again consider $n_1 = 49$ and $n_2 = 48$

EXAMPLE 2

| **Number** | **Difference** | **Base $x$** |
|---|---|---|
| $(x - a)$<br>49 | $a$<br>+1 | 10*5= 50 |
| $(x - b)$<br>48 | $b$<br>+2 | |
| $\{(x - a) + (x - b) - x\} * 5$<br>**235** | $ab$<br>**2** | |

Hence answer = 2352

*Note:*

- The Modified Base Method is applied for all three conditions of $n_1$ and $n_2$ as in *Cases 1.1, 1.2 and 1.3* using either Multiplication or Division shown above.
- If one of the numbers to be multiplied is a Base itself, then we have to consider only the calculation on LHS, as product of differences is zero.
- Number of digits for product $(a \times b)$ equals the power of 10 with which base is associated.

*Extension to Radix-2 Number System:*

The Nikhilam Sūtra as described in [5] is applied to decimal, i.e., Radix – 10 system only. We have extended the domain of *Nikhilam Sūtra* to encompass Binary, i.e., Radix – 2 number system. This enables us to harness it's prowess for implementing the unsigned Multiplication algorithm on digital platform.

All cases discussed for decimal system are applied to hexadecimal in analogous manner and are found to work accurately.

The implementation can also be easily extended to perform signed binary number multiplication. This can be accomplished by processing the MSBs i.e. sign bits of both multiplicands separately. If '1' denotes positive and '0' denotes negative number, then the sign of the Multiplication result can be obtained by simple EX-OR logic applied on the sign bits of the two numbers.

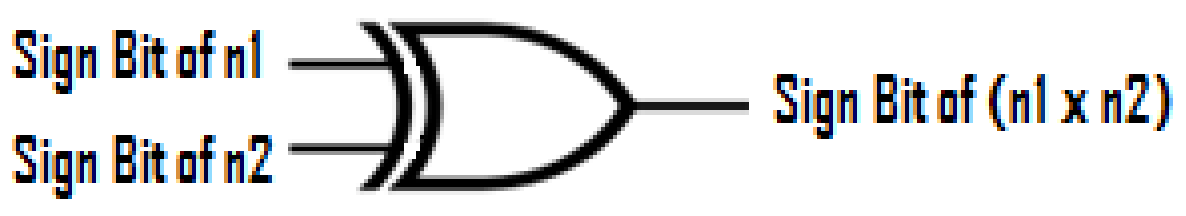

## III. Need For Recursion

*Review of Existing Works*:

The comparison with the existing similar attempts mentioned in references leads to the following observations:

- Our previous work "An Efficient Binary Multiplication Algorithm based on Vedic Mathematics" see [1] is an 8 bit non recursive binary multiplication method. The 8 bit binary multiplication method in [1] is much superior to that in [3]. Even in the worst case the former is faster by over 50% and much faster for the average case.
- The method in [2] compares the Nikhilam Sutra and Urdha Tiryak Sutra and highlights the cases when either of the algorithms is more efficient than the other. Further the paper explains the basic mechanism of Nikhilam Sutra applied to digital arithmetic but is restricted to 4 bit binary multiplication. For higher bit multiplication the algorithm is not extendible easily and needs to consider a very very large number of cases of multiplicands and hence it is unacceptable when compared to [1].
- The 'efficient' multiplication algorithms like Schönhage-Strassen algorithm, Fürer's algorithm only perform well for very large numbers as multiplicands and hence not suitable for general purpose applications.
- The Algorithm in [1] can be extended to higher bit multiplications by breaking up the number into 8 bit segments and operating separately on them however the proposed generalization method is much more efficient than such an implementation.

*Scope for Recursion*:

The Nikhilam Sutra based multiplication can be summarized as the expression of a large multiplication into a small multiplication and an addition. For the Modified Base method, this involves another small multiplication for base adjustment.

The proposed method completely avoids using the Modified Base method by appropriate base selection and hence avoids additional small multiplication. The possible bases in the proposed method are in geometric progression as opposed to arithmetic progression in [1].

Due to such base selection, the differences from Base may not be small as expected by Nikhilam Sutra to be efficient. Thereby in such case, the product of differences may itself be performed by recursively applying the proposed algorithm. This process continues until the final product of differences is very small so as to be accomplished with multiple additions.

In this manner any two numbers, irrespective of their magnitude, can be decomposed successively by the proposed algorithm. Effectively the multiplication is broken down into series of additions, as explained in the following section with examples.

## IV. Recursive Algorithm For Binary Multiplication

We propose a novel algorithm for binary multiplication which depends on the ratio of the number of 1s and the number of 0s in the binary representation of the highest of the two multiplicands.

In this section we first introduce the algorithm, and then explain it with an example. The next section provides a theoretical proof based on the principle of Mathematical Induction.

*Algorithm*:

1. Of the two multiplicands, say X and Y, select the smaller multiplicand.
2. If either X or Y is 0 or 1 then go to step 6.
3. If the second MSB of this smaller multiplicand is 1 then select 'Base' B as 2^b else the select B as 2^(b-1).
4. Subtract this base from both the multiplicands to get differences X' and Y'.
5. Apply steps 1 to 3 to X' and Y' recursively till X' = 1 or Y' = 1 OR X' = 0 or Y' = 0.
6. At recursions, get the partial result at each step by either X + Y' OR Y + X'.
7. Multiply X' and Y' in the final stage (where one of X' or Y' is either 0 or 1.
8. At each step the respective b no. of bits of the partial result are considered while other MSBs are carries generated.
9. Using step 8, merge the partial results to get the final product.

*Example:*

Consider the following example illustrated in binary and verified in the decimal number system. Each column stands for a recursion and shows the partial results obtained at that recursion. The Multiplication result, after merging the partial results is shown in penultimate row, and verification in decimal system is shown in the last row.

$$23 \times 21$$

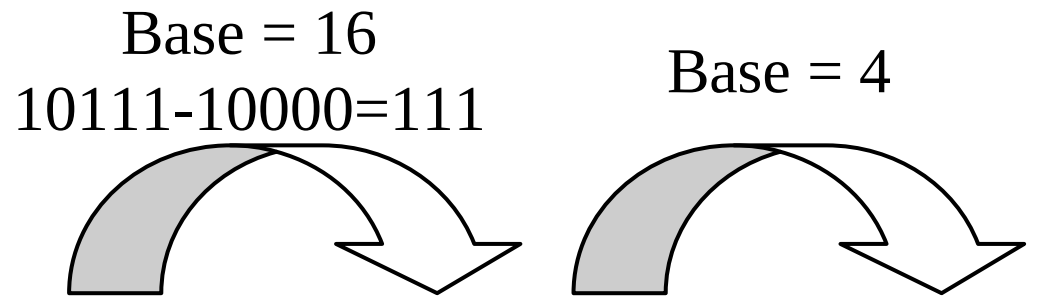

| | | |
|---|---|---|
| 10111 | 111 | 11 |
| 10101 | 101 | 01 |
| 101100<br>11100 | 1000 | 11 |
| | 00 | 11 |
| 10 | | |
| 11110 | | |
| **111100011** | | |
| $23 \times 21 = 483$ | | |

## V. Proof By Mathematical Induction

**Step 1 : For base case…**

$$Base: 2^{b_1}$$

$$\begin{array}{c|c} n_1 & n_1 - 2^{b_1} \\ n_2 & n_2 - 2^{b_1} \end{array}$$

$$n_1 + (n_2 - 2^{b_1}) \qquad (n_1 - 2^{b_1}) \times (n_2 - 2^{b_1})$$

$$= [\, n_1 + (n_2 - 2^{b_1})\,] \times 2^{b_1} + (n_1 - 2^{b_1}) \times (n_2 - 2^{b_1})$$
$$= \quad n_1 \times n_2$$

**Step 2 : Induction Assumption for some 'k'**

Induction assumption for base : $2^{b_k}$

| | | | | |
|---|---|---|---|---|
| $n_1$ | $n_1 - 2^{b_1}$ | $n_1 - 2^{b_1} - 2^{b_2}$ | ... | $n_1 - 2^{b_1} - 2^{b_2} - \dots\dots - 2^{b_k}$ |
| $n_2$ | $n_2 - 2^{b_1}$ | $n_2 - 2^{b_1} - 2^{b_2}$ | ... | $n_2 - 2^{b_1} - 2^{b_2} - \dots\dots - 2^{b_k}$ |
| $\left\{\left[ n_1 + \left( n_2 - 2^{b_1} \right) \right] \times 2^{b_1} \right\}$ | $\left\{\left[ (n_1 - 2^{b_1}) + \left( n_2 - 2^{b_1} - 2^{b_2} \right) \right] \times 2^{b_2} \right\}$ | ……… | ... | $\left[ (n_1 - 2^{b_1} - 2^{b_2} - \dots\dots - 2^{b_k}) \times (n_2 - 2^{b_1} - 2^{b_2} - \dots\dots - 2^{b_k}) \right]$ |

Therefore, Assumption becomes :

$$\left\{\left[\begin{matrix} n_1 + \\ (n_2 - 2^{b_1}) \end{matrix}\right] \times 2^{b_1}\right\} + \left\{\left[\begin{matrix} (n_1 - 2^{b_1}) + \\ (n_2 - 2^{b_1} - 2^{b_2}) \end{matrix}\right] \times 2^{b_2}\right\} + \dots\dots\dots$$
$$+ \left[\begin{matrix} (n_1 - 2^{b_1} - 2^{b_2} - \dots\dots - 2^{b_k}) \times \\ (n_2 - 2^{b_1} - 2^{b_2} - \dots\dots - 2^{b_k}) \end{matrix}\right] = n_1 \times n_2$$

i.e.

$$\left\{\left[\begin{matrix} n_1 + \\ (n_2 - 2^{b_1}) \end{matrix}\right] \times 2^{b_1}\right\} + \left\{\left[\begin{matrix} (n_1 - 2^{b_1}) + \\ (n_2 - 2^{b_1} - 2^{b_2}) \end{matrix}\right] \times 2^{b_2}\right\} + \dots\dots\dots$$
$$= (n_1 \times n_2) - \left[\begin{matrix} (n_1 - 2^{b_1} - 2^{b_2} - \dots\dots - 2^{b_k}) \times \\ (n_2 - 2^{b_1} - 2^{b_2} - \dots\dots - 2^{b_k}) \end{matrix}\right] \quad \text{…..Equation (1)}$$

**Step 3 : Proof for base $2^{b_{k+1}}$ using assumption in previous case**

| | | | |
|---|---|---|---|
| $n_1$ | ... | $n_1 - 2^{b_1} - 2^{b_2} - \dots \dots - 2^{b_k}$ | $n_1 - 2^{b_1} - 2^{b_2} - \dots\dots - 2^{b_k} - 2^{b_{k+1}}$ |
| $n_2$ | ... | $n_2 - 2^{b_1} - 2^{b_2} - \dots\dots - 2^{b_k}$ | $n_2 - 2^{b_1} - 2^{b_2} - \dots\dots - 2^{b_k} - 2^{b_{k+1}}$ |
| $\left\{\left[ n_1 + \left( n_2 - 2^{b_1} \right) \right] \times 2^{b_1} \right\}$ | ... | $\left\{\left[ (n_1 - 2^{b_1} - 2^{b_2} - \dots\dots - 2^{b_k}) + (n_2 - 2^{b_1} - 2^{b_2} - \dots\dots - 2^{b_k} - 2^{b_{k+1}}) \right] \times 2^{b_{k+1}} \right\}$ | $\left[ (n_1 - 2^{b_1} - 2^{b_2} - \dots\dots - 2^{b_k} - 2^{b_{k+1}}) \times (n_2 - 2^{b_1} - 2^{b_2} - \dots\dots - 2^{b_k} - 2^{b_{k+1}}) \right]$ |

$$Let \quad x = n_1 - 2^{b_1} - 2^{b_2} - \dots\dots - 2^{b_k}$$
$$and \quad y = n_2 - 2^{b_1} - 2^{b_2} - \dots\dots - 2^{b_k}$$

Therefore above expression gives us,

$$\left\{\left[\begin{matrix} n_1 + \\ (n_2 - 2^{b_1}) \end{matrix}\right] \times 2^{b_1}\right\} + \dots\dots\dots + \left\{\left[\begin{matrix} x + \\ (y - 2^{b_{k+1}}) \end{matrix}\right] \times 2^{b_{k+1}}\right\}$$
$$+ [(x - 2^{b_{k+1}}) \times (y - 2^{b_{k+1}})]$$

$$= \left\{\left[\begin{matrix} n_1 + \\ (n_2 - 2^{b_1}) \end{matrix}\right] \times 2^{b_1}\right\} + \dots\dots\dots + \left\{\left[\begin{matrix} x + \\ (y - 2^{b_{k+1}}) \end{matrix}\right] \times 2^{b_{k+1}}\right\}$$
$$+ [(x - 2^{b_{k+1}}) \times (y - 2^{b_{k+1}})]$$

Using result from equation (1) we get :

$$= (n_1 \times n_2) - \left[\begin{matrix} (n_1 - 2^{b_1} - 2^{b_2} - \dots\dots - 2^{b_k}) \times \\ (n_2 - 2^{b_1} - 2^{b_2} - \dots\dots - 2^{b_k}) \end{matrix}\right]$$
$$+ \left\{\left[\begin{matrix} x + \\ (y - 2^{b_{k+1}}) \end{matrix}\right] \times 2^{b_{k+1}}\right\}$$
$$+ [(x - 2^{b_{k+1}}) \times (y - 2^{b_{k+1}})]$$

$$= (n_1 \times n_2) - [x \times y] + \left\{\left[\begin{matrix} x + \\ (y - 2^{b_{k+1}}) \end{matrix}\right] \times 2^{b_{k+1}}\right\}$$
$$+ [(x - 2^{b_{k+1}}) \times (y - 2^{b_{k+1}})]$$

$$= (n_1 \times n_2) - [x \times y]$$
$$+ \{ x \times 2^{b_{k+1}} + y \times 2^{b_{k+1}} - 2^{2b_{k+1}} \}$$
$$+ [\, x \times y - x \times 2^{b_{k+1}} - y \times 2^{b_{k+1}} + 2^{2b_{k+1}} \,]$$
$$= n_1 \times n_2 - x \times y + x \times 2^{b_{k+1}} + y \times 2^{b_{k+1}} - 2^{2b_{k+1}}$$
$$+ x \times y - x \times 2^{b_{k+1}} - y \times 2^{b_{k+1}}$$
$$+ 2^{2b_{k+1}}$$
$$= n_1 \times n_2$$

Thus above Induction proves the correctness of the algorithm and it holds for all bases.

## VI. Hardware Implementation

*Determining the base:*

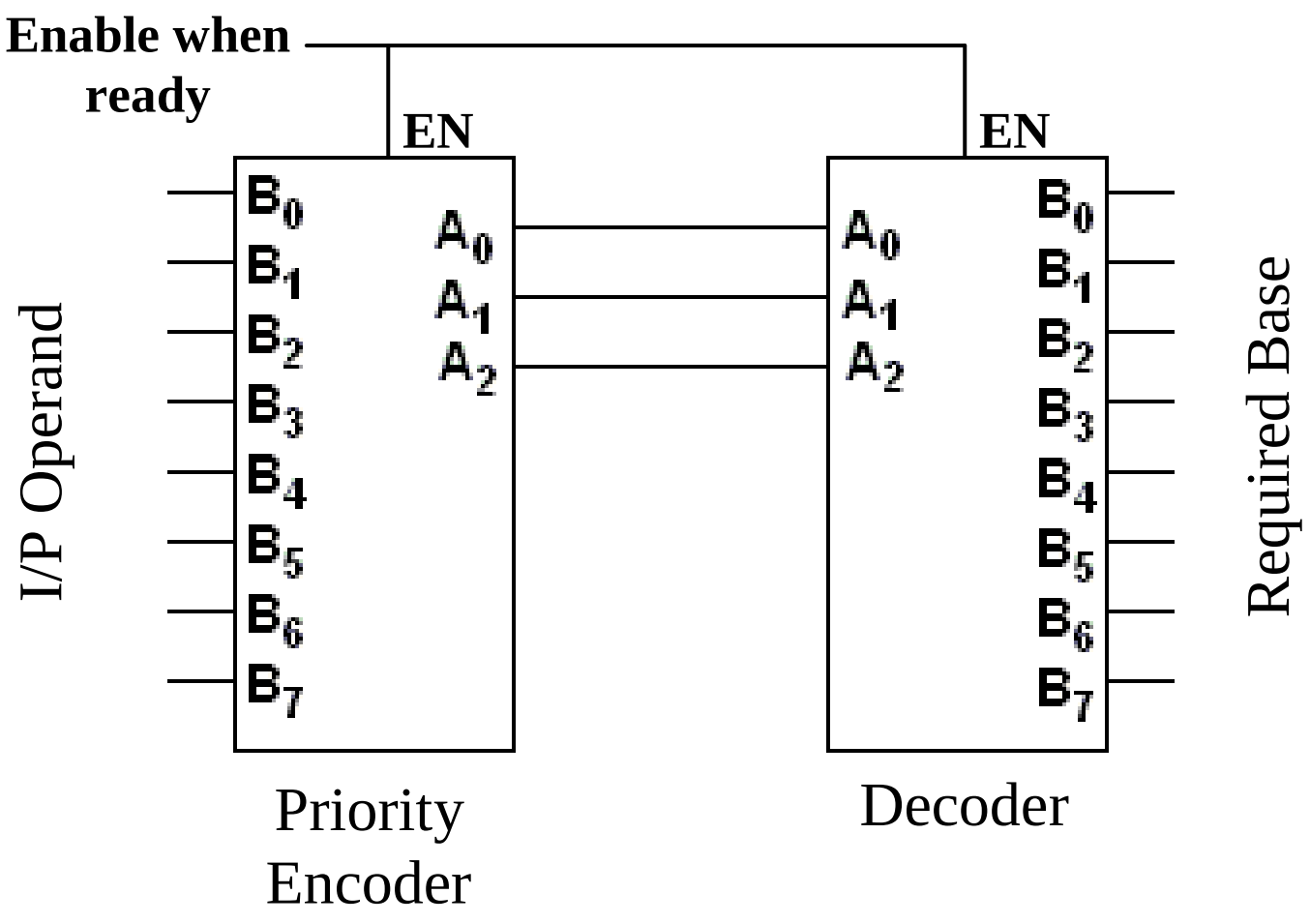

The base calculator allows determination of required base in one clock cycle.

- Consider the input operand 00110110
- The priority encoder output will be $A_2A_1A_0 = 101$
- Effectively the decoder will output the required base viz. 00100000

*Multiplier Hardware:*

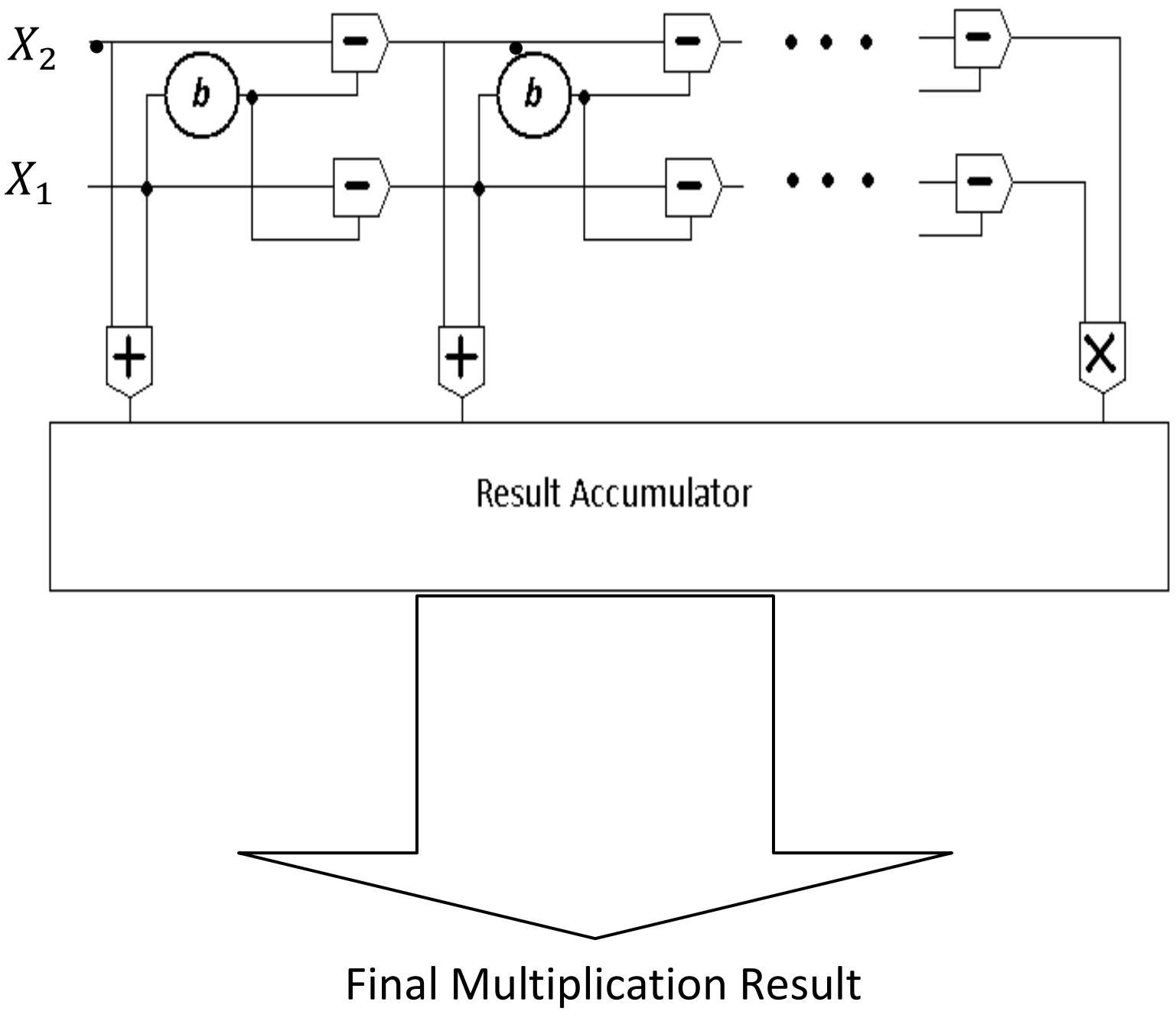

where,

= base calculator

## VII. Statistical Analysis

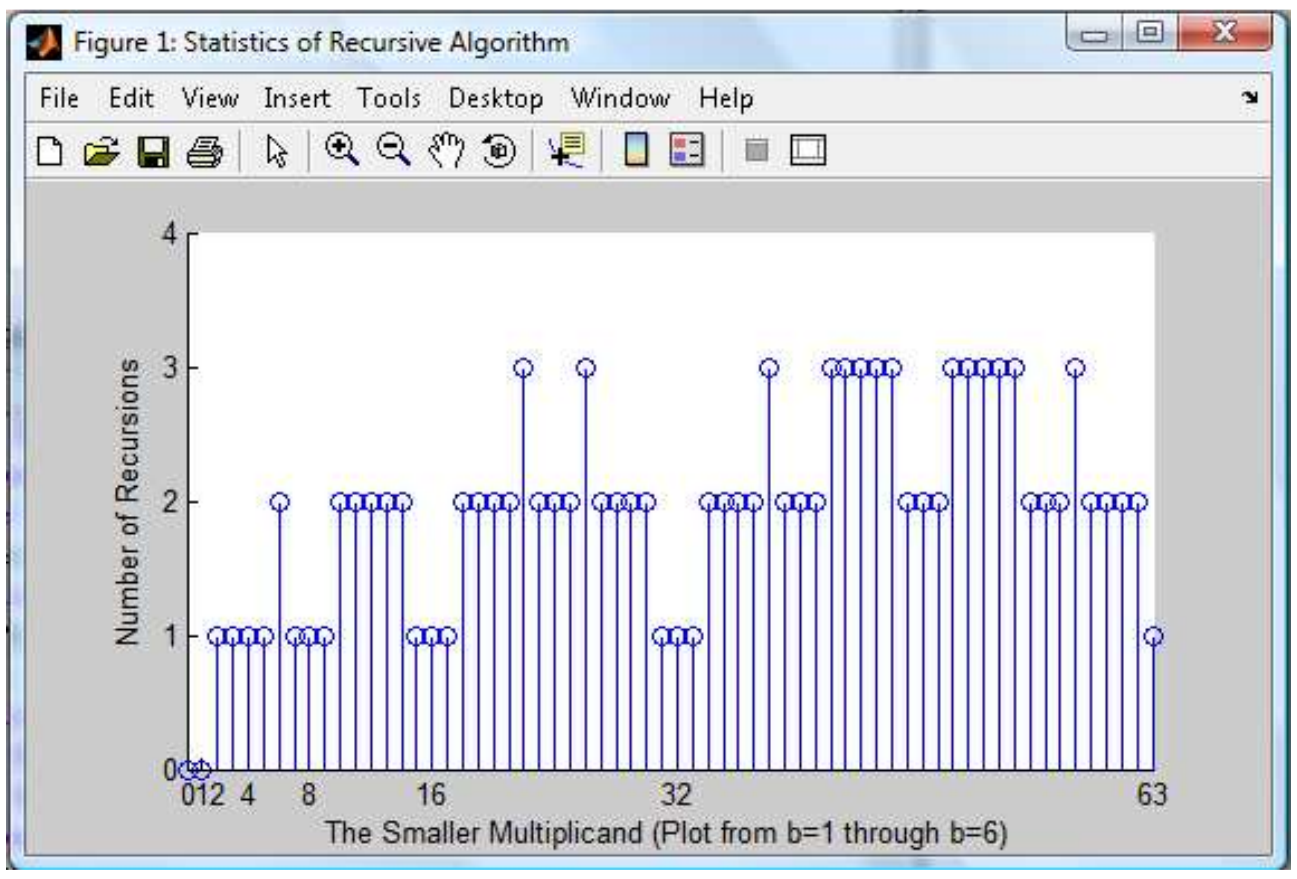

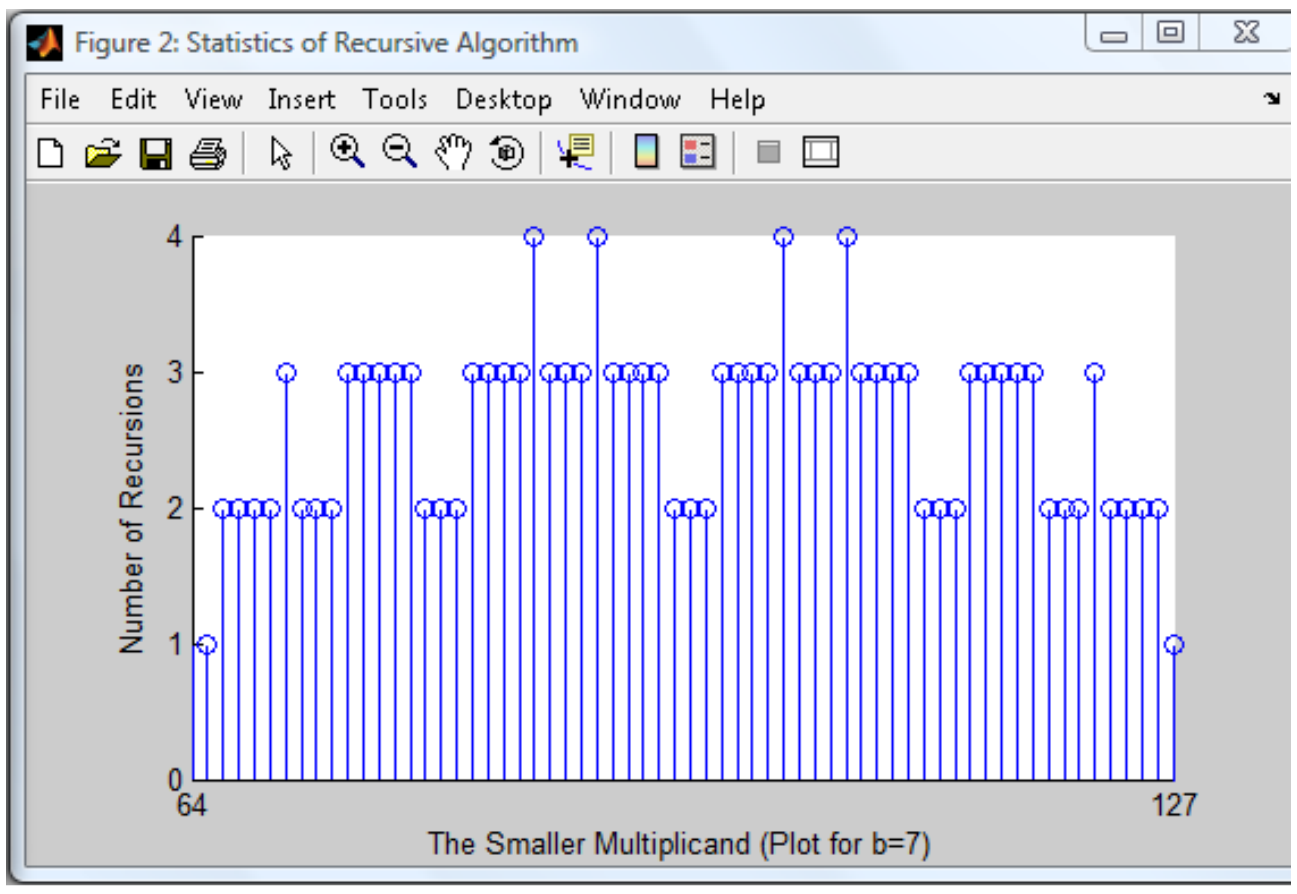

The plots describe the number of recursions required for each multiplicand from 0 to 127, where that multiplicand is the smaller of the two multiplicands. X-axis represents the smaller multiplicand ‘n’ and Y-axis shows corresponding number of recursions. Intervals shown on x-axis are the points where the number of bits ‘b’ changes, where ‘b’ is the number of bits needed to represent ‘n’.

It can be easily seen form the plots that there is a pattern in the recursions for every ‘b’ bit number. The information about the number of occurrences of worst case i.e. number of maximum recursions cases for any multiplication involving smaller multiplicand of ‘b’ bits can be seen from the plots.

If f(b) represents this count of cases with maximum number of recursions required for any multiplication involving smaller multiplicand of ‘b’ bits, then we have the recurrence relation for f(b) from the plots as follows.

*Initial Values:*

f(1) = 0, f(2) = 0, f(3) = 1, f(4) = 5

*Recurrence Relations:*

If ‘b’ is odd and b>3
f(b) = 2*f(b-2)
If ‘b’ is even and b>4
f(b) = 2*f(b-2) + f(b-1)

From the above relations, f(b) can be calculated for any ‘b’ which gives us quantitative idea about number of worst case occurrences for each ‘b’.

The equations show that for odd values of ‘b’, the cardinality of worst case recursions increases by a fixed multiple of 2 over that of ‘b-2’ where as the corpus and the range increase as a power of 2. For the case where 'b' is even valued, the cardinality of worst case recursions increases only linearly depending on cardinalities for ‘b-1’ and ‘b-2’, whereas the range still increases in powers of 2.

The worst case recursions being ceil(b/2) the worst case complexity of the algorithm will be O(n*b).

## VII. Conclusion And Future Work

We presented an algorithm which can be used in future to compute multiplication of two variable bit numbers. The algorithm was found to solely depend on the ratio of the number of 1's and 0's used to represent a number in binary, rather than on the magnitude of the number. It was seen from the graphs that as the ratio approaches 1 the number of operations required for the multiplication increases and decreases as the ratio tends to move close to 0. The statistical analysis provided a clear idea of the efficiency of the algorithm over the range of input multiplicands. The algorithm was found to be implementable by the hardware proposed and is much more suitable for applications on multiprocessor environments, than those proposed in [1], [2] and [3].

In future we plan to work on the average case analysis of the algorithm we propose. We intend to do the hardware simulation of the algorithm to analyze the empirical results of the multiplications over a corpus of numbers.